\documentclass[12pt]{article}
\usepackage[latin1]{inputenc}
\usepackage[authoryear, nonamebreak, elide]{natbib}
\usepackage{graphicx}
\usepackage{amssymb, amsmath, amsthm}
\usepackage{times}
\usepackage{color}
\usepackage{bm}
\usepackage{multirow}
\usepackage{subfig}
\usepackage{rotating}
\usepackage{lscape}
\newcommand{\vecbeta}{\boldsymbol{\beta}}

\newcommand{\vecx}{\boldsymbol{x}}
\newcommand{\vecX}{\boldsymbol{X}}
\newcommand{\vecescore}{\boldsymbol{U}}

\newcommand\Top{\rule{0pt}{3.0ex}}       % Top strut 
\newcommand\B{\rule[-1.2ex]{0pt}{0pt}} % Bottom strut

\begin{document}

\title{Log-symmetric models with cure fraction with application to leprosy reactions  data}
\author{
%\normalsize
\textbf{\normalsize Joyce B. Rocha}$^{1}$,\, \textbf{\normalsize Francisco M. C. Medeiros}$^{1}$ \textbf{\normalsize and} \textbf{\normalsize Dione M. Valen{\c{c}}a}$^{1}$\\[-0.1cm]
{\footnotesize  Department of Statistics, Federal University of Rio Grande do Norte, Natal, Brazil.}\\[-0.15cm]
}
\date{}
%\title{Log-symmetric models with cure fraction with application to leprosy reactions  data}
%
%
%
\maketitle
%\centerline{\bf \today}
%\date{}

\begin{abstract}
\noindent
In this paper, we propose a log-symmetric survival model with cure fraction,  considering that the  distributions of lifetimes for susceptible individuals belong to the log-symmetric class of distributions. This class has continuous, strictly positive, and asymmetric distributions, including the log-normal, log-$t$-Student, Birnbaum-Saunders, log-logistic I, log-logistic II, log-normal-contaminated, log-exponential-power, and log-slash distributions. The log-symmetric class is quite flexible and allows for including bimodal distributions and outliers. This includes explanatory variables  through the parameter associated with the cure fraction.  We evaluate the performance of the proposed model through  extensive simulation studies and consider a real data application to evaluate the effect of factors on the immunity to leprosy reactions in patients with Hansen's disease.

\noindent {\it Keywords:}  Cure rate. Log-symmetric models. Maximum likelihood. Survival analysis.
\end{abstract}

%%%%%%%%%%%%%%%%%%%%%%%%%%%%%%%%%%%%%%%%%%%%%%%%%%%%%%%%%%%%%%%%%%%%%%%%%%%%%                                                                                                                  %%%%%%%%%%%%%%%%%%%%%%%%%%%%%%%%%%Introduction%%%%%%%%%%%%%%%%%%%%%%%%%%%%%%%
%%%%%%%%%%%%%%%%%%%%%%%%%%%%%%%%%%%%%%%%%%%%%%%%%%%%%%%%%%%%%%%%%%%%%%%%%%%%%

\section{Introduction}\label{sec:intro}
\noindent
		
In the parametric approach of survival analysis, some probabilistic models such as exponential, Weibull, log-normal, and log-logistic are  often used to fit lifetime data in practical situations (see  \cite{Lawless11}, \cite{KP2002}, and \cite{CO84} for applications and inferential properties). Several proposals for generalized and extended distributions have been presented to provide more flexibility in modeling lifetime data. We can cite the generalized gamma \citep{stacy62}, F-generalized \citep{peng98}, generalized inverse Gaussian \citep{Jorgensen82}, and generalized modified Weibull \citep{carrasco08} distributions, among others. \cite{Lai13} describes some common methods for constructing lifetime distributions. %because they present a good fit which are distributions with several known models as particular case (s,

However, there are several distributions in the literature defined on the positive real line that can fit  survival data. Here, the interest lies in the log-symmetric class of distributions, which is obtained from an exponential transformation of a random variable with symmetric distribution and characterized by continuous, strictly positive, and asymmetric distributions  (see \cite{vanegas16}). According to \cite{medeiros17}, this class includes distributions with lighter and heavier tails than  log-normal distributions, such as the log-normal, log-logistic, log-$t$-Student, Harmonic law, Birnbaum-Saunders, Birnbaum-Saunders-$t$, Birnbaum-Saunders generalized, log-normal-contaminated, log-exponential-power, and log-slash. \cite{vanegas16} studied the statistical properties of this class and verified that the two parameters are interpreted directly as location and scale. The scale parameter is the dispersion of the data and the  location  is the median, which is a robust measure in the presence of outliers and informative in survival analysis. In fact,  according to \cite{Lawless11}, for lifetime distributions, the median is more used than the mean since it is easier to estimate when the data are censored and always exists for proper distributions, while the mean may not exist.

%The location parameter is the median, which is a robust measure in the presence of outliers, and the scale is the dispersion of the data. In survival analysis the median is a very informative measure. According 
% (since $ S (x) <0.5 $, when $ x $ tends to infinity)

\cite{vanegas2017log} studied log-symmetric models to fit survival data. They proposed a semi-parametric regression model to analyze strictly asymmetric data in the presence of non-informative censoring. The authors used a nonlinear structure for the median and a non-parametric structure to model the asymmetry (dispersion) parameter considering  models of the log-symmetric class. These models relax the assumption of log-normal errors, including other distributions of this class.

A  basic assumption of classic  survival analysis is that all individuals will present the event of interest if they are followed for a sufficient time. However, in some situations, this is not true since some individuals could not be susceptible to the event,  even for a long follow-up. These individuals are called immune or cured \citep{maller1996}. Survival models that  deal with these situations are known as long-term models or cure rate models. The best-known long-term models are  the  standard mixture model, introduced by \cite{boag1949} and \cite{berkson1952survival}, and the promotion time model (also known as bounded cumulative hazards), proposed by \cite{yakovlev1993} and extended by \cite{chen1999new}. 

\cite{rodrigues2009unification} proposed a unified long-term model, in which not only the distribution of the times until the occurrence of the event of interest (also called latency distribution) may take different forms, but also the distribution of the number of competing causes for the occurrence of the event of interest (also known as incidence distribution). In this approach, the standard mixture and promotion time models represent particular cases in which the incidence distribution assumes, respectively,  Bernoulli and Poisson distributions. \cite{ortega2009} presented the promotion time model with the generalized log-gamma distribution for the latency. \cite{Canchoelouzada2012} presented a cure rate % long-term
model assuming the geometric distribution  for the incidence and the Birnbaum-Saunders distribution for the latency. \cite{fonseca2013} presented a simulation study considering missing covariates in the promotion time model, with Weibull as the distribution of latency. \cite{Hashimoto2014} proposed a survival promotion time model in which the latency follows a Birnbaum-Saunders distribution. 
In the present paper, we propose the log-symmetric model with cure fraction. We consider that the latency follows a member of the log-symmetric class of distributions, and for the incidence we study the   Bernoulli, Poisson and geometric distributions.  Thus, results in \cite{Hashimoto2014} and \cite{Canchoelouzada2012} follow as special cases of the more general results given here.% in this paper.
%Thus, we generalizes the results of the models proposed by . 

This paper is organized as follows. Section \ref{MRLS-sec:2} defines the log-symmetric model with cure rate. In Section \ref{MRLS-sec:3}, we obtain  the likelihood and score functions for the general model and some particular cases. In Section \ref{MRLS-sec:4}, we evaluate the performance of the proposed model through an extensive Monte Carlo simulation study. In Section \ref{MRLS-sec:5}, we present and discuss an empirical application on patients with leprosy to evaluate the effect of factors on immunity to leprosy reactions and show the applicability of the proposed model. Section \ref{MRLS-sec:6} closes the paper with final remarks. %Technical details are presented in an appendix. 

%%%%%%%%%%%%%%%%%%%%%%%%%%%%%%%%%%%%%%%%%%%%%%%%%%%%%%%%%%%%%%%%%%%%%%%%%%%%%                             %%%%%%%%%%Symmetric linear regression models, estimation and testing%%%%%%%%%
%%%%%%%%%%%%%%%%%%%%%%%%%%%%%%%%%%%%%%%%%%%%%%%%%%%%%%%%%%%%%%%%%%%%%%%%%%%%%

\section{The model}\label{MRLS-sec:2}
\noindent

For an individual $i$  in the population, let $M_i$  be a latent variable denoting the number of  causes or risks competing for the occurrence of the event of interest with probability function $p_{\theta}(m_i)=P_{\theta}(M_i=m_i)$ (incidence distribution). The time-to-event (for the $i$-th  individual) due to $j$-th  cause  is denoted by $Z_{ij}$, $i = 1,\ldots, n$ and $ j = 1, \ldots, M_i$. Given  $M_i=m_i $, we assume that $Z_{i1}, Z_{i2}, \ldots, Z_{im_i}$ are independent and identically distributed with  common probability density function (latency distribution) given by
\begin{align}\label{dls}
f(z; \eta, \phi)=\frac{g(\tilde{z}^2)}{z\sqrt{\phi}}, \qquad z>0,
\end{align}
where $\tilde{z} = \log\left[(z/\eta)^\frac{1}{\sqrt{\phi}}\right]$, $\eta>0$ is {the median of $Z_{ij}$, and $\phi>0$ is a shape (skewness or relative dispersion) parameter, for some function $g:\mathbb{R}\rightarrow [0,\infty)$ (called density generating function)  such that $\int_{0}^{\infty}u^{-1/2}g(u)du=1$. We write $Z_{ij}\sim \text{LS}(\eta,\phi^2, g(\cdot))$ and denote its common  survival function by $S(z; \eta, \phi)$.}

{This class of distributions is called log-symmetric because $W=\log(Z_{ij})$ belongs to the symmetric class of distributions with parameters $\mu=\log(\eta)$ and $\phi$, density generating function $g(\cdot)$, and probability density function given by 
$f_W(w;\mu,\phi)=g((w-\mu)^2/\phi)/\sqrt{\phi}, w\in\mathbb{R}.$ 
In particular, a symmetric distribution with $ \mu = 0 $ and $ \phi = 1$ is called standard symmetric distribution with probability density and distribution functions represented here, respectively, by % a symmetric distribution, with $\mu= 0$ and $\phi=1$ have the standard symetric distribution, with  distribution functions denoted here by
$f_0(w)=g(w^2)$ and $F_0(w)=\int^{w}_{-\infty}g(u^2)du$.} 
%Log-symmetric distributions are studied in \cite{vanegas16} and provides a wide range of distributions to model continuous positive data. 
Different choices for the density generating function $g(\cdot)$ lead to different distributions in \eqref{dls}. See Table \ref{tab1:1} for some examples.

Let $T_i$ be a random variable representing the time-to-event defined as $T_i= \min\{Z_{i0}, Z_{i1}, \ldots, Z_{iM_i}\}$, where the sequence $Z_{i0}, Z_{i1}, \ldots$ does not depend on $ M_i$ and $ P (Z_{i0} = \infty) = 1 $. This assumption permits the occurrence of immune individuals (\textit{infinite lifetimes}) since $ M_i = 0$ means that there are no causes or risks for the occurrence of the event. Under this setup, the long-term survival function, the sub-density function, and the sub-hazard rate function for $T_i$  are given, respectively, by
\begin{eqnarray}\label{eq:Sp}
	S_p(t) &=& P(T_i > t) = p_{\theta}(0) + \sum_{m=1}^{\infty} p_{\theta}(m)S_0(\tilde{t})^{m},\nonumber \\ 
	f_p(t) &=& -\dfrac{\partial S_{p}(t)}{\partial t} = \frac{g(\tilde{t}^2)}{t\sqrt{\phi}} \sum_{m=0}^{\infty} m p_{\theta}(m)S_0(\tilde{t})^{m-1}, \\
	h_p(t) &=&  \dfrac{f_{p}(t)}{S_{p}(t)}= \frac{g(\tilde{t}^2)\sum_{m=0}^{\infty} m p_{\theta}(m)S_0(\tilde{t})^{m-1}}{t\sqrt{\phi}\sum_{m=0}^{\infty} p_{\theta}(m)S_0(\tilde{t})^{m}},\nonumber
\end{eqnarray}
where $S_0(\cdot)=1-F_0(\cdot)$ is the survival function of the standard symmetric distribution and $\tilde{t} = \log\left[(t/\eta)^\frac{1}{\sqrt{\phi}}\right]$. Hence, $S_p(t)$ is an improper survival function since $\displaystyle{\lim_{t\rightarrow\infty}}S_p(t) = p_{\theta}(0)>0$, where $p_{\theta}(0)$ represents the cure fraction (proportion of cured or immune individuals) in the population. Below, we present a few specific models that arise from our general formulation. Particularly, we consider situations where $M_i$ has Bernoulli, Poisson, and geometric distributions. %The functions $f_p(t)$ and $h_p(t)$ are improper functions. 

\begin{table}[!h]
%\footnotesize
{  \small
\centering
\caption{Density generating function for some log-symmetric distributions.$^a$}\label{tab1:1}
%\resizebox{\linewidth}{!}
{ 
\begin{tabular}{llll}
\hline
Distribution && &$g(u), \quad u>0$\Top\B \\
\hline
log-normal    &&  & $\dfrac{\exp(-u/2)}{\sqrt{2\pi}}$\Top\\                
\\
log-$t$-Student        &&  & $\dfrac{\nu^{-1/2}}{B(1/2,\nu/2)}\left(1+\dfrac{u}{\nu}\right)^{\dfrac{-(\nu+1)}{2}}, \quad \nu \in \mathbb{R}$  \\  &&  & \\
Birnbaum-Saunders       &&  & $\dfrac{1}{\sqrt{2\pi}}\exp\left(\dfrac{-2}{\alpha^2}\sinh^2[\sqrt{u}]\right)\dfrac{2}{\alpha}\cosh(\sqrt{u}),$ $\quad \alpha>0$\\ 
\\
type I log-logistic         &&  & $c\dfrac{e^{-u}}{(1+e^{-u})^2}, \quad c \cong 1.4843$\\ 
\\
type II log-logistic        &&  & $\dfrac{e^{-\sqrt{u}}}{(1+e^{-\sqrt{u}})^2}$\\ 
\\
log-power-exponential       &&  & $\dfrac{1}{\Gamma(1+\frac{1+k}{2})2^{1+(1+k)/2}}\exp\left(\dfrac{-1}{2}u^{1/(1+k)}\right)$, $\quad -1< k \leq1$ \B\\
\hline
\multicolumn{4}{l}{{\small $^a$ {$B(\cdot, \cdot)$ and $\Gamma(\cdot)$ are the beta and gamma functions, respectively.}}}

%$C(k)=\Gamma(1+\frac{1+k}{2})2^{1+(1+k)/2}$									
%\multicolumn{4}{l}{{\small $^a$$\Gamma(\cdot)$ is the gamma function.}}
\end{tabular} \\	
}
}
\end{table}

\vspace{0.5cm}
\textbf{1. Log-symmetric standard mixture model}: If $M_i$ follows a Bernoulli distribution with $p_{\theta}(1)=(1-\theta)$ ($0<\theta<1$), we obtain the classical mixture model \citep{boag1949, berkson1952survival}, {where the proportion of cured individuals in the population is given by $\theta=p_{\theta}(0)$}. The long-term survival function, the sub-density, and sub-hazard rate functions for $T_i$  are, respectively,  % The density function and survival function are, respectively,
\begin{eqnarray}\label{LSSMP}
	S_p(t) &=& \theta+(1-\theta)S_0(\tilde{t}), \nonumber \\
	f_p(t) &=&  \frac{1}{t\sqrt{\phi}}(1-\theta)g(\tilde{t}^2),  \\
	h_p(t) &=&  \frac{(1-\theta)g(\tilde{t}^2)}{t\sqrt{\phi}[\theta+(1-\theta)S_0(\tilde{t})]}. \nonumber
	\end{eqnarray}	
	
\textbf{2. Log-symmetric promotion time model:} If $M_i$ follows a Poisson distribution with mean $\theta>0$, we obtain the model proposed by \cite{chen1999new} {with cure fraction given by $\exp(-\theta)=p_{\theta}(0)$}. The long-term survival function, the sub-density, and sub-hazard rate functions for $T_i$  are, respectively, given by
\begin{eqnarray}\label{LSPTM}
S_p(t) &=&  \exp[-\theta F_0(\tilde{t})], \nonumber\\
	f_p(t) &=&  \frac{\theta g(\tilde{t}^2)\exp[-\theta F_0(\tilde{t})]}{t\sqrt{\phi}}, \\
	h_p(t) &=&  \frac{\theta g(\tilde{t}^2)\exp[-\theta F_0(\tilde{t})]}{t\sqrt{\phi}\exp[-\theta F_0(\tilde{t})]}. \nonumber
\end{eqnarray}

\textbf{3. Log-symmetric geometric model:} If $M_i$ follows a geometric distribution with probability function $p_{\theta}(m)=\theta(1-\theta)^m$, where $0<\theta<1$, the long-term survival function, the sub-density function, and the sub-hazard rate function are defined, respectively, by
\begin{eqnarray}\label{LSGM}
S_p(t) &=&  \frac{\theta}{1-(1-\theta)S_0(\tilde{t})}, \nonumber\\
	f_p(t) &=&  \frac{\theta(1-\theta)g(\tilde{t}^2)}{t\sqrt{\phi}[1-(1-\theta)S_0(\tilde{t})]^2}, \\
	h_p(t) &=&  \frac{(1-\theta)g(\tilde{t}^2)}{t\sqrt{\phi}[1-(1-\theta)S_0(\tilde{t})]}.\nonumber
	\end{eqnarray}
The cure fraction is given by $\theta=p_{\theta}(0)$.
%\end{center}

%%%%%%%%%%%%%%%%%%%%%%%%%%%%%%%%%%%%%%%%%%%%%%%%%%%%%%%%%%%%%%%%%%%%%%%%%%%%%
%%%%%%%%%%%%%%%%%%%%%%%%%%%%%Inference%%%%%%%%%%%%%%%%%%%%%%%%%%%%%%%%%%%%%%%
%%%%%%%%%%%%%%%%%%%%%%%%%%%%%%%%%%%%%%%%%%%%%%%%%%%%%%%%%%%%%%%%%%%%%%%%%%%%%

\section{Inference}\label{MRLS-sec:3}
\noindent

Consider that the time-to-event may not always be observed, being subject to a right censoring time (random and non-informative). For each individual $i$, $i = 1,\ldots, n$, denote by  $C_i$ the censoring time variable and let $Y_i =\min\{T_i, C_i\}$ be the observable lifetime, where $C_i$ is independent of $T_i$. Let $\delta_i$ be the failure/censoring indicator, with $\delta_i = 1$ if $T_i \leq C_i$ and $\delta_i = 0$ if $T_i > C_i$.

We incorporate covariates in the parametric cure rate model through the relation {$\theta_i = q(\bm{x_i}^\top; \vecbeta)$}, where $\vecx_i=(x_{i0},x_{i1}\ldots, x_{ip})^\top$   is the vector of covariates associated to the $i$-th observation ($x_{i0}=1, \forall i$), $\vecbeta=(\beta_{0},\beta_{1}\ldots, \beta_{p})^\top$ is the vector of unknown parameters, and $q(\cdot)$ is a continuous, invertible, and twice differentiable function, called the link function, which links the covariates $\bm{x_i}$ to the parameter of interest $\theta_i$. Note that when covariates are included in the model, we have different cure rate parameters, $\theta_i$,  $i = 1,\ldots, n$, for each individual. We assume that $\vecX=(\vecx_1,\cdots,\vecx_n)^\top$ is a full-rank $n\times p$ matrix, i.e. $\mathrm{rank}(\vecX)=p$, and that usual regularity conditions for likelihood inference are valid \citep[Chap.9]{CoxHinkley1974}. To simplify the notation, {consider} the $n$-dimensional vectors of observations $\bm{y} = (y_1, y_2, \ldots, y_n)^\top$, $\bm{\delta} = (\delta_1, \delta_2, \ldots, \delta_n)^\top$, and $\bm{m} = (m_1,m_2,\ldots ,m_n)^\top$. Hence, the complete dataset is denoted by $D_c = (n, \bm{y}, \bm{\delta},\bm{m}, \bm{X})$, and the dataset without the latent variables is denoted by $D = (n, \bm{y}, \bm{\delta}, \bm{X})$.
In the {standard mixture and geometric models, the most used relation to associate the parameter $\theta_i$ with the covariates is the logistic link function \citep{maller1996} given by}
\begin{eqnarray}
\theta_i = \frac{\exp(\bm{x}^{\top}_i\bm{\beta})}{1+\exp(\bm{x}^{\top}_i\bm{\beta})}.\nonumber
\end{eqnarray}
In the promotion time model, the relation often used to associate the parameter $\theta_i$ with the covariates is given by the logarithmic link function \citep{chen1999new}, expressed by	
\begin{eqnarray}
\theta_i = \exp(\bm{x}^{\top}_i\bm{\beta}).\nonumber
\end{eqnarray}

Thus, the vector of unknown parameters in the model is denoted by $\bm{\lambda} = (\vecbeta^\top, \eta, \phi)^\top$, and after some algebra, it can be shown that the log-likelihood function for the complete data $D_c$ is given by
%f(y_i; \eta, \phi)
\begin{align}\label{logDc}
\ell(\bm{\lambda} ;\mathcal{D}_c) = \sum^{n}_{i=1}\delta_i \log m_i+\sum^{n}_{i=1}m_i\log 
S(y_i; \eta, \phi)+\sum^{n}_{i=1}\delta_i\log\frac{f(y_i; \eta, \phi)}{S(y_i; \eta, \phi)} + \sum^{n}_{i=1}\log p_{\theta_i}(m_i). 
\end{align}

Note that the likelihood (\ref{logDc}) is not observable since it depends on the latent variables. The marginal likelihood for the observed data is obtained by summing over all possible values for the variables $M_i$,
$i = 1,\ldots, n$.

Therefore, the logarithm of the marginal likelihood function is given by
\begin{eqnarray}\label{logM}
\ell(\bm{\lambda};\mathcal{D}) = \sum_{i=1}^{n}\delta_i\log f_p(y_i;\bm{\lambda})+\sum_{i=1}^{n}(1-\delta_i)\log S_p(y_i;\bm{\lambda}),
\end{eqnarray}
where in a regression context associated with the incidence model, $f_p(y_i;\bm{\lambda})$ and $S_p(y_i;\bm{\lambda})$ are obtained in \eqref{eq:Sp} by replacing $ \theta $  by $ \theta_i = q(\bm {x_i} ^ \top; \bm{\beta})$. The use of marginal likelihood in cure rate models is common 
(see for example \citealp{tsodikov1998, Cancho2011, mizoi2007, rodrigues2009unification, ortega2009, fonseca2013, loose2018}).  In addition to the marginal likelihood being considered an ordinary  likelihood \citep{Cox1975}, an additional attraction for using this approach is that \eqref{logM} appears to be a generalization of the usual (log) likelihood considered in survival models with the presence of censoring. The demonstration of \eqref{logM} can be found in \cite{carneiro2016}.

%\dione{where $f_p(y_i;\bm{\lambda})$ and $S_p(y_i;\bm{\lambda})$ are equivalent to (\ref{eq:Sp}), replacing $ \theta $  by $ \theta_i = q(\bm {x_i} ^ \top; \vecbeta) $, in the incidence model}
The score vector for $\bm{\lambda} = (\vecbeta^\top, \eta, \phi)^\top$  is given by $\vecescore(\bm{\lambda}) = (\vecescore_{\vecbeta}(\bm{\lambda})^{\top}, U_{\eta}(\bm{\lambda}), U_{\phi}(\bm{\lambda}))^{\top}$, \linebreak
where $\vecescore_{\vecbeta}(\bm{\lambda})=
\partial\ell(\bm{\lambda};\mathcal{D})/
\partial\bm{\beta}=(U_{\beta_1}(\bm{\lambda}), 
\ldots, U_{\beta_p}(\bm{\lambda}))^{\top}_{p\times1}$, $U_{\eta}(\bm{\lambda})={\partial\ell(\bm{\lambda};\mathcal{D})}/{\partial\eta}$, and $U_{\phi}(\bm{\lambda})={\partial\ell(\bm{\lambda};\mathcal{D})}/{\partial\phi}$.

%\partial \ell(\bm{\lambda};\mathcal{D})/\partial \bm{beta} =(U_{\beta_1}(\bm{\lambda})$, %U_{\beta_2}(\bm{\lambda}),\ldots, U_{\beta_p}(\bm{\lambda}))^{\top}$ 
The maximum likelihood estimate %of $\bm{\lambda} = (\vecbeta^\top, \eta, \phi)^\top$,
$\widehat{\bm{\lambda}} = (\widehat{\vecbeta}^\top, \widehat{\eta}, \widehat{\phi})^\top$ is obtained by simultaneously solving the nonlinear equations $\vecescore_{\vecbeta}(\bm{\lambda})={\bf 0}$, $U_{\eta}(\bm{\lambda})=0$, and $U_{\phi}(\bm{\lambda})=0$. This system of equations cannot be analytically solved and statistical software can be {used} to solve it numerically. In general, in the presence of censored observations, the expected Fisher information matrix cannot be obtained. Thus, inferences are based on the observed information matrix. Asymptotically,
\[
(\widehat{\vecbeta}^\top, \widehat{\eta}, \widehat{\phi})^\top \sim N_{p+2}\left((\vecbeta^\top, \eta, \phi)^\top, {\ddot{{\cal{L}}}(\bm{\lambda})}^{-1} \right), 
\]
where $\ddot{{\cal{L}}}(\bm{\lambda})=-\partial^2 \ell(\bm{\lambda};\mathcal{D})/\partial \bm{\lambda}\bm{\lambda}^\top$ is the $(p + 2)\times(p + 2)$ observed information matrix.   
  
%We now give the log-likelihood function and the score vector some log-symmetric models with fraction cure.
Next, we present the log-likelihood function and score vector for $\bm{\lambda} = (\vecbeta^\top, \eta, \phi)^\top$ considering models \eqref{LSSMP}, \eqref{LSPTM}, and \eqref{LSGM}.
\vspace{0.5cm}

\textbf{1. Log-symmetric standard mixture model} 

\begin{itemize}
    \item[\textit{i)}]  Marginal log-likelihood function:
\begin{align*}
\ell(\bm{\lambda};\mathcal{D})&=\sum^{n}_{i=1} \delta_i\left[\log(1-\theta_i)+\log g(\tilde{y}_i^2)-\log(y_i\sqrt{\phi})\right]\\ 
       &+ \sum^{n}_{i=1}(1-\delta_i)\log\left[\theta_i+(1-\theta_i)(1- F_0(\tilde{y}_i))\right]. \nonumber
\end{align*}

\item[\textit{ii)}] Components of the score vector:% for $\vecbeta$, $\eta$, and $\phi$ %is given, respectively, by %$\vecescore(\bm{\lambda}) = (\vecescore_{\vecbeta}(\bm{\lambda})^{\top}, U_{\eta}(\bm{\lambda}), U_{\phi}(\bm{\lambda}))^{\top}$ with
\begin{align*}
U_{\beta_l}(\bm{\lambda}) &= \sum_{i=1}^{n} -\delta_i\theta_ix_{il} + \sum_{i=1}^{n}\frac{(1-\delta_i)F_0(\tilde{y}_i)\theta_i(1-\theta_i)}{\theta_i+(1-\theta_i)(1-F_0(\tilde{y}_i))}x_{il}, \mbox{ for } l=0,1,\dots,p,\\
U_{\eta}(\bm{\lambda}) &= \sum_{i=1}^{n} \frac{\delta_i}{g(\tilde{y}^2_i)}\frac{\partial g(\tilde{y}^2_i)}{\partial\eta} + \frac{1}{\eta\sqrt{\phi}}\sum_{i=1}^{n}\frac{(1-\delta_i)(1-\theta_i)f_0(\tilde{y}_i)}{\theta_i+(1-\theta_i)(1-F_0(\tilde{y}_i))]},\\
U_{\phi}(\bm{\lambda})&=\sum_{i=1}^{n} \frac{\delta_i}{g(\tilde{y}^2_i)}\frac{\partial g(\tilde{y}^2_i)}{\partial\phi}+\frac{1}{2\phi}\sum_{i=1}^{n}\frac{(1-\delta_i)(1-\theta_i)f_0(\tilde{y}_i)\tilde{y}_i}{\theta_i+(1-\theta_i)(1-F_0(\tilde{y}_i))}.
\end{align*}
\end{itemize}
\vspace{0.5cm}
%\textbf{2. 
\textbf{2. Log-symmetric promotion time model}

\begin{itemize}
    \item[\textit{i)}]  Marginal log-likelihood function:
\begin{align*}
	\ell(\bm{\lambda};\mathcal{D})=\sum^{n}_{i=1} \delta_i\left[\log(\theta_i)+\log g(\tilde{y}_i^2)-\theta_iF_0(\tilde{y}_i)-\log(y_i\sqrt{\phi})\right]-\sum^{n}_{i=1}(1-\delta_i)\theta_iF_0(\tilde{y}_i).
\end{align*}
%and the components of score vector for $\vecbeta$, $\eta$, and $\phi$ are, respectively,
\item[\textit{ii)}] Components of the score vector:
\begin{align*}
U_{\beta_l}(\bm{\lambda}) &= \sum_{i=1}^{n} \delta_ix_{il}- \sum_{i=1}^{n}F_0(\tilde{y}_i)\theta_ix_{il},\mbox{ for } l=0,1,\dots,p,\\
U_{\eta}(\bm{\lambda}) &= \sum_{i=1}^{n} \frac{\delta_i}{g(\tilde{y}^2_i)}\frac{\partial g(\tilde{y}^2_i)}{\partial\eta} - \frac{1}{\eta\sqrt{\phi}}\sum_{i=1}^{n}\theta_if_0(\tilde{y}_i),\\
U_{\phi}(\bm{\lambda}) &= \sum_{i=1}^{n} \frac{\delta_i}{g(\tilde{y}^2_i)}\frac{\partial g(\tilde{y}^2_i)}{\partial\phi} -\frac{1}{2\phi}\sum_{i=1}^{n}\delta_i +\frac{1}{2\phi}\sum_{i=1}^{n}\theta_if_0(\tilde{y}_i)\tilde{y}_i.
\end{align*}
\end{itemize}

%\textbf{3. 
\textbf{3. Log-symmetric geometric model} 
\begin{itemize}
    \item[\textit{i)}]  Marginal log-likelihood function:
\begin{align*}
	\ell(\bm{\lambda};\mathcal{D})&=\sum^{n}_{i=1} \delta_i\left[\log(\theta_i)+\log(1-\theta_i)+\log g(\tilde{y_i}^2)-\log(y_i\sqrt{\phi})\right] \nonumber\\
& + \sum^{n}_{i=1} \delta_i[-2\log(\theta_i+(1-\theta_i)F_0(\tilde{y}_i))] \nonumber\\
&+ \sum^{n}_{i=1} (1-\delta_i)\left[\log(\theta_i)-\log(\theta_i+(1-\theta_i)F_0(\tilde{y}_i))\right].
\end{align*}

\item[\textit{ii)}] Components of the score vector (for $ l=0,1,\dots,p$):
\begin{align*}
U_{\beta_l}(\bm{\lambda}) &= \sum_{i=1}^{n} \delta_i(1-2\theta_i)x_{il}+\sum_{i=1}^{n}(1-\delta_i)(1-\theta_i)x_{il}-\sum_{i=1}^{n}\frac{(1+\delta_i)(1-F_0(\tilde{y}_i))\theta_i(1-\theta_i)}{\theta_i+(1-\theta_i)F_0(\tilde{y}_i)}x_{il},\\
U_{\eta}(\bm{\lambda}) &= \sum_{i=1}^{n} \frac{\delta_i}{g(\tilde{y}^2_i)}\frac{\partial g(\tilde{y}^2_i)}{\partial\eta} + \frac{1}{\eta\sqrt{\phi}}\sum_{i=1}^{n}\frac{(1+\delta_i)(1-\theta_i)f_0(\tilde{y}_i)}{\theta_i+(1-\theta_i)F_0(\tilde{y}_i)},\\
U_{\phi}(\bm{\lambda}) &=  \sum_{i=1}^{n} \frac{\delta_i}{g(\tilde{y}^2_i)}\frac{\partial g(\tilde{y}^2_i)}{\partial\phi}-\frac{1}{2\phi}\sum_{i=1}^{n}\delta_i + \frac{1}{2\phi}\sum_{i=1}^{n}\frac{(1+\delta_i)(1-\theta_i)f_0(\tilde{y}_i)\tilde{y}_i}{\theta_i+(1-\theta_i)F_0(\tilde{y}_i)}.
\end{align*}
\end{itemize}

%%%%%%%%%%%%%%%%%%%%%%%%%%%%%%%%%%%%%%%%%%%%%%%%%%%%%%%%%%%%%%%%%%%%%%%%%%%%%%%%%%%%%%%%%%%%%%%%%%%%%Simulation results%%%%%%%%%%%%%%%%%%%%%%%%%%%%%%%%%%%
%%%%%%%%%%%%%%%%%%%%%%%%%%%%%%%%%%%%%%%%%%%%%%%%%%%%%%%%%%%%%%%%%%%%%%%%%%%%%

\section{Simulation results}\label{MRLS-sec:4}
\noindent

In this section, we shall present a Monte Carlo simulation study to investigate and compare the performance of the maximum likelihood estimators in log-symmetric promotion time cure models. We considered the following latency distributions: log-normal, log-$t$-Student with $\nu = 3$ degrees of freedom, and Birnbaum-Saunders extended with $\alpha=1.5$. The values for $x_{i1}$ and $x_{i3}$ were obtained as random draws from a uniform distribution in the interval $(0, 1)$, and the values for $x_{i2}$ were randomly obtained from the Bernoulli distribution with a success probability of $0.5$. The censoring times $C_i$ were generated as independent random variables uniformily distributed in the interval $[0, u]$, where $u$ was suitably chosen to produce the following censoring percentages: $15\%$ and $30\%$.

To define the proportion of censoring used in the simulation, we considered the approach given in \cite{fonseca2013} and the following relation:
\[
cp_{total} = cp (1-cf) + cf,
\]
where $ cp $ is the censoring proportion among those susceptible to the event,  $ cp_{total} $ is the censoring proportion in relation to all units under study (susceptible or cured)  and $ cf $  is the cure fraction.
Although in real data applications $cp_{total}$ is the only calculable measure, we also considered the censoring among uncured ($cp$) since this allows us to distinguish between censored and cured individuals in the simulation study.

Three different sample sizes were considered: $n= 250$,  $n=500$, and $n=1000$. For the $i$-th cured individual, $M_i$ was generated as a Poisson distribution with mean $\theta_i = \exp(\bm{x}_i^\top \vecbeta)$, representing the incidence distribution,  % number of risk possibilities for the $i$-th individual,
$\vecbeta = (\beta_0, \beta_1, \beta_2, \beta_3)^\top$. Different cure fractions in the sample were  obtained  by  changing  the  value  of $\vecbeta$. Thus, $\vecbeta=(0.42,0.25,0.24,0.34)^\top$ leads to $cf =10\%$ and  $\vecbeta= (0.10,0.05,0.07,0.03)$ leads to $cf =30\%$. The median and shape parameter are $\eta=5$ and $\phi=1$, respectively.

The number of Monte Carlo replicates was 5000 and all simulations were performed in the R software \citep{R2020}. All the parameters, except the assumed known parameters  $\nu$ (log-$t$-Student) and $\alpha$ (Birnbaum-Saunders extended), were estimated by the maximum likelihood method. {The optimizations were performed} using the quasi-Newton method BFGS (Broyden-Fletcher-Goldfarb-Shanno) through the function \textit{optim}. The evaluation of the point estimation was carried out based on the following quantities for each sample size: {mean}, relative bias, the root of the relative mean square error, and standard error, which are given, respectively, by
\begin{align*}
\begin{split}
	&\mbox{mean}(\widehat{\gamma})= \frac{1}{5000} \sum_{r=1}^{5000} \widehat{\gamma}_r, \hspace{2cm} \mbox{RB}= \frac{\mbox{mean}(\widehat{\gamma})-\gamma}{\gamma}, \\ 
	&\sqrt{\mbox{RMSE}}= \frac{1}{5000} \sum_{r=1}^{5000} \left(\frac{\widehat{\gamma}_r - \gamma}{\gamma}\right)^2 \hspace{0.5cm}
	\mbox{se} = \sqrt{\frac{\displaystyle{\sum_{r=1}^{5000}} (\widehat{\gamma_r}-\mbox{mean}(\widehat{\gamma}))^2}{5000 - 1}},
\end{split}
\end{align*}
where $\widehat{\gamma}$ is the parameter estimate of the $r$-th replicate. The results are presented in Tables 2 and 3 (log-normal); Tables 4 and 5 (log-$t$-Student); and Tables 6 and 7 (Birnbaum-Saunders extended).
%Tables \ref{tab:LNbetas}-\ref{tab:LTetaphi}

The results suggest that the estimates of parameters $\beta_1$, $\beta_2$, and $\beta_3$ are close to their true values, even with high censure proportion.  For instance,  for $n = 500$, $cp=30\%$, and $cf= 30\%$, the estimates of the parameters for the log-normal promotion time model (Table \ref{tab:LNbetas}) are $0.051$, $0.072$, and $0.031$, for the log-$t$-Student promotion time model (Table \ref{tab:LTbetas}) we obtain $0.054$, $0.076$, and $0.031$, respectively, and for the Birnbaum-Saunders promotion time model (Table \ref{tab:BSbetas}) we have $0.053$, $0.073$, and $0.030$.%. The same behavior was noted for the estimates of the parameters $\eta$ and $\phi$ for all the models considered, see Tables 3, 5, and 7, respectively.

We note that the relative bias of $\beta_0$, $\eta$, and $\phi$ increases as the censoring proportion and cure fractions increase. For example, consider the log-$t$-Student promotion time model  with $n=250$ in Tables \ref{tab:LTbetas} and \ref{tab:LTetaphi}, the relative bias of $\widehat{\beta}_0$ increases from $-0.145$  ($cp = 15\%$ and $cf=10\%$) to $0.600$ ($cp =30 \%$ and $cf=30\%$), and the relative bias of $\widehat{\eta}$ ($\widehat{\phi}$) increases from $-0.010$ ($0.028$)  (with $cp = 15\%$ and $cf=10\%$) to $0.338$ ($0.195$) (with $cp =30 \%$ and $cf=30\%$). Comparing the results presented in these tables, we observe that as the sample size increases, in general, the bias of the estimators reduces, as expected.

%%%%%%%%%%%%%%%%%%%%%%%%%%%%%%%%%%%%%%%%%%%%%%%%%%%%%%%%%%%%%%%%%%%%%%%%%%%%%%%%%%%%%%%%%%%%%%%%%%%%%%%%%%%MODELO LOG NORMAL%%%%%%%%%%%%%%%%%%%%%%%%%%%%%%
%%%%%%%%%%%%%%%%%%%%%%%%%%%%%%%%%%%%%%%%%%%%%%%%%%%%%%%%%%%%%%%%%%%%%%%%%%%%

%Estimates of regression coefficients for different values of n, censoring percentage (cp), and cure fraction (cf):log-normal promotion time model

\begin{table}[htbp!]
\caption{Log-normal promotion time model: Estimates of the regression coefficients  for different values of $n$, censoring percentage ($cp$), and cure fraction ($cf$), with true values of $\vecbeta$ for each  $cf$ $(\vecbeta_{cf})$ given by   $\vecbeta_{10}= (0.42,0.25,0.24,0.34)^\top$ and $\vecbeta_{30} =(0.10,0.05,0.07,0.03)^\top$.}
\resizebox{\linewidth}{!}{ 
		\begin{tabular}{cccccccccccccccccccccc}
			\hline
\multirow{2}{*}{\centering $n$} & \multirow{2}{*}{\centering $cp$} & \multirow{2}{*}{\centering $cf$}& \multicolumn{4}{c}{$\widehat{\beta}_0$}& &\multicolumn{4}{c}{$\widehat{\beta}_1$}& & \multicolumn{4}{c}{$\widehat{\beta}_2$}& &\multicolumn{4}{c}{$\widehat{\beta}_3$}\Top\B\\
 \cline{4-7} \cline{9-12} \cline{14-17} \cline{19-22}   
 					  & & 	& mean &  RB & $\sqrt{\mbox{RSME}}$ & se & & mean &  RB & $\sqrt{\mbox{RSME}}$ & se & & mean &  RB & $\sqrt{\mbox{RSME}}$ & se && mean &  RB & $\sqrt{\mbox{RSME}}$ & se \Top\B\\
\hline 
\multirow{4}{*}{\centering 250}
 & 
 \multirow{2}{*}{\centering $15$}
	 & $10$ & $0.362 $&$ -0.138 $&$ 0.235 $&$ 0.228 $& &$ 0.265 $&$ 0.060 $&$ 0.257 $&$ 0.256 $&&$ 0.251 $&$ 0.046 $&$ 0.150 $&$ 0.149 $&&$ 0.366 $&$ 0.076 $&$ 0.271 $&$ 0.270$ \Top\\ 
	  &  & $30$ & $0.011 $&$ -0.890 $&$ 0.255 $&$ 0.239 $& &$ 0.055 $&$ 0.100 $&$ 0.291 $&$ 0.291 $& &$ 0.070 $&$ 0.000 $&$ 0.167 $&$ 0.167 $& &$ 0.027 $&$ -0.100 $&$ 0.302 $&$ 0.302$ \\ 
 %&  &  &  &  & &  &  &  &  &  &  &  &  &  &  &  & & \\ 
 %\cline{2-19}	  
&                        
  \multirow{2}{*}{\centering $30$}
	  & $10$ & $0.565 $&$ 0.345 $&$ 0.408 $&$ 0.382 $ & &$ 0.268 $&$ 0.072 $&$ 0.283 $&$ 0.283 $&&$ 0.254 $&$ 0.058 $&$ 0.166 $&$ 0.165 $ &&$ 0.372 $&$ 0.094 $&$ 0.298 $&$ 0.296$  \\ 
	  &  & $30$ & $0.223 $&$ 1.230 $&$ 0.408 $&$ 0.389 $&& $ 0.055 $&$ 0.100 $&$ 0.322 $&$ 0.322 $&&$ 0.072 $&$ 0.029 $&$ 0.185 $&$ 0.185 $& &$ 0.031 $&$ 0.033 $&$ 0.331 $&$ 0.331$ \\  
  &  &  &  &  & &  &  &  &  &  &  &  &  &  &  &  \\
\multirow{4}{*}{\centering 500}
 &
 \multirow{2}{*}{\centering $15$}
	 & $10$ & $0.356 $&$ -0.152 $&$ 0.178 $&$ 0.166 $&&$ 0.264 $&$ 0.056 $&$ 0.178 $&$ 0.177 $&&$ 0.250 $&$ 0.042 $&$ 0.105 $&$ 0.104 $&&$ 0.354 $&$ 0.041 $&$ 0.189 $&$ 0.189$ \\ 
	  &  & $30$ & $0.009 $&$ -0.910 $&$ 0.198 $&$ 0.176 $&&$ 0.051 $&$ 0.020 $&$ 0.203 $&$ 0.203 $&& $ 0.070 $&$ 0.000 $&$ 0.117 $&$ 0.117 $&&$ 0.031 $&$ 0.033 $&$ 0.210 $&$ 0.210$ \\ 
 %&  &  &  &  & &  &  &  &  &  &  &  &  &  &  &  & & \\ 
 %\cline{2-19}	  
&                              
  \multirow{2}{*}{\centering $30$}
	  & $10$ & $0.537 $&$ 0.279 $&$ 0.288 $&$ 0.263 $&&$ 0.268 $&$ 0.072 $&$ 0.196 $&$ 0.195 $&&$ 0.253 $&$ 0.054 $&$ 0.117 $&$ 0.116 $&&$ 0.361 $&$ 0.062 $&$ 0.207 $&$ 0.206$ \\ 
	  &  & $30$ &  $0.197 $&$ 0.970 $&$ 0.281 $&$ 0.264 $&&$ 0.051 $&$ 0.020 $&$ 0.224 $&$ 0.224 $&&$ 0.072 $&$ 0.029 $&$ 0.130 $&$ 0.130 $&&$ 0.031 $&$ 0.033 $&$ 0.232 $&$ 0.232$ \\ 
    &  &  &  &  &  &  &  &  &  &  &  &  &  &  &  &  \\
\multirow{4}{*}{\centering 1000}
 &
 \multirow{2}{*}{\centering $15$}
	  & $10$ &  $0.354 $&$ -0.157 $&$ 0.133 $&$ 0.116 $&&$ 0.261 $&$ 0.044 $&$ 0.126 $&$ 0.126 $&&$ 0.249 $&$ 0.038 $&$ 0.074 $&$ 0.073 $&&$ 0.358 $&$ 0.053 $&$ 0.131 $&$ 0.129$\\ 
	  &  & $30$ &  $0.008 $&$ -0.920 $&$ 0.153 $&$ 0.123 $&&$ 0.051 $&$ 0.020 $&$ 0.142 $&$ 0.142 $&&$ 0.071 $&$ 0.014 $&$ 0.082 $&$ 0.082 $&&$ 0.029 $&$ -0.033 $&$ 0.147 $&$ 0.147$ \\ 
 %&  &  &  &  & &  &  &  &  &  &  &  &  &  &  &  & & \\ 
 %\cline{2-19}	  
&                              
  \multirow{2}{*}{\centering $30$}
	  & $10$ & $0.521 $&$ 0.240 $&$ 0.206 $&$ 0.180 $&&$ 0.265 $&$ 0.060 $&$ 0.140 $&$ 0.139 $&&$ 0.253 $&$ 0.054 $&$ 0.081 $&$ 0.081 $&&$ 0.360 $&$ 0.059 $&$ 0.143 $&$ 0.142$ \\ 
	  &  & $30$ & $0.188 $&$ 0.880 $&$ 0.203 $&$ 0.183 $&&$ 0.053 $&$ 0.060 $&$ 0.158 $&$ 0.158 $&&$ 0.073 $&$ 0.043 $&$ 0.090 $&$ 0.090 $&&$ 0.030 $&$ 0.000 $&$ 0.160 $&$ 0.160$ \B\\ 
   \hline
		\end{tabular}
}
\label{tab:LNbetas}
\end{table}

\begin{table}[htbp!]
\centering
\caption{Log-normal promotion time model: Estimates of $\eta$ and $\phi$ (with true values  $\eta = 5$ and $\phi = 1$).}
%\resizebox{23cm}{!}{
%\resizebox{\linewidth}{!}{ 
		\begin{tabular}{cccccccccccc}
			\hline
\multirow{2}{*}{\centering $n$} & \multirow{2}{*}{\centering $cp$} & \multirow{2}{*}{\centering $cf$} & \multicolumn{4}{c}{$\widehat{\eta}$}& &\multicolumn{4}{c}{$\widehat{\phi}$}\Top\B\\
 \cline{4-7} \cline{9-12}
 			& & & mean &  RB & $\sqrt{\mbox{RSME}}$ & se && mean &  RB & $\sqrt{\mbox{RSME}}$ & se \Top\B \\
\hline 
\multirow{4}{*}{\centering 250}
 & 
 \multirow{2}{*}{\centering $15$}
	  & $10$    & $5.067 $&$ 0.013 $&$ 0.932 $&$ 0.930 $& &$ 1.020 $&$ 0.020 $&$ 0.174 $&$ 0.173$ \Top\\ 
	  &  & $30$ & $4.842 $&$ -0.032 $&$ 0.714 $&$ 0.696 $& &$ 1.010 $&$ 0.010 $&$ 0.176 $&$ 0.176$ \\ 
 %&  &  &  &  & &  &  &  &  &  &  &  &  &  &  &  & & \\ 
 %\cline{2-19}	  
&                        
  \multirow{2}{*}{\centering $30$}
	  & $10$    & $7.190 $&$ 0.438 $&$ 5.676 $&$ 5.237 $& &$ 1.201 $&$ 0.201 $&$ 0.384 $&$ 0.327$ \\ 
	  &  & $30$ & $7.861 $&$ 0.572 $&$ 14.798 $&$ 14.521 $& &$ 1.312 $&$ 0.312 $&$ 0.521 $&$ 0.418$ \\ 
   &  &  &  &  &  &  &  &  &  &    \\
\multirow{4}{*}{\centering 500}
 &
 \multirow{2}{*}{\centering $15$}
	  & $10$    & $4.972 $&$ -0.006 $&$ 0.598 $&$ 0.597 $& &$ 1.015 $&$ 0.015 $&$ 0.122 $&$ 0.121$ \\ 
	  &  & $30$ & $4.785 $&$ -0.043 $&$ 0.508 $&$ 0.460 $& &$ 1.007 $&$ 0.007 $&$ 0.122 $&$ 0.122$  \\   
 %\cline{2-19}	  
&                              
  \multirow{2}{*}{\centering $30$}
	  & $10$    & $6.482 $&$ 0.296 $&$ 2.586 $&$ 2.119 $& &$ 1.179 $&$ 0.179 $&$ 0.282 $&$ 0.218$ \\ 
	  &  & $30$ & $6.752 $&$ 0.350 $&$ 2.938 $&$ 2.359 $& &$ 1.281 $&$ 0.281 $&$ 0.388 $&$ 0.268$\\  
   &  &  &  &  &  &  &  &  &  &   \\
\multirow{4}{*}{\centering 1000}
 &
 \multirow{2}{*}{\centering $15$}
	  & $10$    & $4.928 $&$ -0.014 $&$ 0.412 $&$ 0.406 $& &$ 1.014 $&$ 0.014 $&$ 0.084 $&$ 0.083$ \\ 
	  &  & $30$ & $4.749 $&$ -0.050 $&$ 0.405 $&$ 0.318 $& &$ 1.002 $&$ 0.002 $&$ 0.087 $&$ 0.087$ \\  

 %\cline{2-19}	  
&                              
  \multirow{2}{*}{\centering $30$}
	  & $10$    & $6.192 $&$ 0.238 $&$ 1.682 $&$ 1.187 $& &$ 1.166 $&$ 0.166 $&$ 0.222 $&$ 0.147$  \\ 
	  &  & $30$ & $6.489 $&$ 0.298 $&$ 2.007 $&$ 1.347 $& &$ 1.269 $&$ 0.269 $&$ 0.326 $&$ 0.184$ \B\\ 
   \hline
\end{tabular}
\label{tab:LNetaphi}
\end{table}

%%%%%%%%%%%%%%%%%%%%%%%%%%%%%%%%%%%%%%%%%%%%%%%%%%%%%%%%%%%%%%%%%%%%%%%%%%%%%%%%%%%%%%%%%%%%%%%%%%%%%
%                            MODELO LOG T
%%%%%%%%%%%%%%%%%%%%%%%%%%%%%%%%%%%%%%%%%%%%%%%%%%%%%%%%%%%%%%%%%%%%%%%%%%%%%%%%%%%%%%%%%%%%%%%%%%%%%

\begin{table}[htbp!]
\caption{Log-$t$-Student promotion time model: Estimates of the regression coefficients  for different values of $n$, censoring percentage ($cp$), and cure fraction ($cf$), with true values of $\vecbeta$ for each  $cf$ $(\vecbeta_{cf})$ given by   $\vecbeta_{10}= (0.42,0.25,0.24,0.34)^\top$ and $\vecbeta_{30} =(0.10,0.05,0.07,0.03)^\top$. }
\resizebox{\linewidth}{!}{ 
		\begin{tabular}{cccccccccccccccccccccc}
			\hline
\multirow{2}{*}{\centering $n$} & \multirow{2}{*}{\centering $cp$} & \multirow{2}{*}{\centering $cf$}& \multicolumn{4}{c}{$\widehat{\beta}_0$}& &\multicolumn{4}{c}{$\widehat{\beta}_1$}& & \multicolumn{4}{c}{$\widehat{\beta}_2$}& &\multicolumn{4}{c}{$\widehat{\beta}_3$}\Top\B\\
 \cline{4-7} \cline{9-12} \cline{14-17} \cline{19-22}   
 					  & & 	& mean &  RB & $\sqrt{\mbox{RSME}}$ & se & & mean &  RB & $\sqrt{\mbox{RSME}}$ & se & & mean &  RB & $\sqrt{\mbox{RSME}}$ & se && mean &  RB & $\sqrt{\mbox{RSME}}$ & se \Top\B\\
\hline 
\multirow{4}{*}{\centering 250}
 & 
 \multirow{2}{*}{\centering $15$}
	  & $10$ & $0.359 $&$ -0.145 $&$ 0.227 $&$ 0.219 $& &$ 0.260 $&$ 0.040 $&$ 0.256 $&$ 0.256 $& &$ 0.252 $&$ 0.050 $&$ 0.152 $&$ 0.152 $& &$ 0.352 $&$ 0.035 $&$ 0.263 $&$ 0.263$  \Top\\ 
	  &  & $30$ & $0.014 $&$ -0.860 $&$ 0.247 $&$ 0.231 $& &$ 0.057 $&$ 0.140 $&$ 0.278 $&$ 0.278 $& &$ 0.071 $&$ 0.014 $&$ 0.168 $&$ 0.168 $& &$ 0.031 $&$ 0.033 $&$ 0.304 $&$ 0.304$   \\ 
 %&  &  &  &  & &  &  &  &  &  &  &  &  &  &  &  & & \\ 
 %\cline{2-19}	  
&                        
  \multirow{2}{*}{\centering $30$}
	  & $10$ & $0.541 $&$ 0.288 $&$ 0.379 $&$ 0.359 $& &$ 0.263 $&$ 0.052 $&$ 0.284 $&$ 0.283 $& &$ 0.253 $&$ 0.054 $&$ 0.168 $&$ 0.167 $& &$ 0.354 $&$ 0.041 $&$ 0.289 $&$ 0.289$  \\ 
	  &  & $30$ & $0.160 $&$ 0.600 $&$ 0.344 $&$ 0.339 $& &$ 0.057 $&$ 0.140 $&$ 0.310 $&$ 0.310 $&&$ 0.072 $&$ 0.029 $&$ 0.184 $&$ 0.184 $& &$ 0.031 $&$ 0.033 $&$ 0.336 $&$ 0.336$  \\ 
  &  &  &  &  & &  &  &  &  &  &  &  &  &  &  &  \\
\multirow{4}{*}{\centering 500}
 &
 \multirow{2}{*}{\centering $15$}
	  & $10$ & $0.351 $&$ -0.164 $&$ 0.175 $&$ 0.161 $& &$ 0.260 $&$ 0.040 $&$ 0.179 $&$ 0.179 $& &$ 0.249 $&$ 0.038 $&$ 0.101 $&$ 0.101 $& &$ 0.357 $&$ 0.050 $&$ 0.186 $&$ 0.185$ \\ 
	  &  & $30$ & $0.014 $&$ -0.860 $&$ 0.191 $&$ 0.171 $& &$ 0.051 $&$ 0.020 $&$ 0.200 $&$ 0.200 $& &$ 0.074 $&$ 0.057 $&$ 0.117 $&$ 0.117 $& &$ 0.030 $&$ 0.000 $&$ 0.206 $&$ 0.206$   \\ 
 %&  &  &  &  & &  &  &  &  &  &  &  &  &  &  &  & & \\ 
 %\cline{2-19}	  
&                              
  \multirow{2}{*}{\centering $30$}
	  & $10$ & $0.514 $&$ 0.224 $&$ 0.255 $&$ 0.238 $& &$ 0.265 $&$ 0.060 $&$ 0.199 $&$ 0.198 $& &$ 0.250 $&$ 0.042 $&$ 0.112 $&$ 0.112 $& &$ 0.359 $&$ 0.056 $&$ 0.205 $&$ 0.204$  \\ 
	  &  & $30$ & $0.133 $&$ 0.330 $&$ 0.237 $&$ 0.235 $& &$ 0.054 $&$ 0.080 $&$ 0.223 $&$ 0.223 $&&$ 0.076 $&$ 0.086 $&$ 0.130 $&$ 0.130 $& &$ 0.031 $&$ 0.033 $&$ 0.230 $&$ 0.230$  \\ 
    &  &  &  &  & &  &  &  &  &  &  &  &  &  &  &  \\
\multirow{4}{*}{\centering 1000}
 &
 \multirow{2}{*}{\centering $15$}
	  & $10$ & $0.353 $&$ -0.160 $&$ 0.133 $&$ 0.115 $& &$ 0.258 $&$ 0.032 $&$ 0.129 $&$ 0.129 $& &$ 0.247 $&$ 0.029 $&$ 0.075 $&$ 0.074 $& &$ 0.351 $&$ 0.032 $&$ 0.129 $&$ 0.129$  \\ 
	  &  & $30$ & $0.012 $&$ -0.880 $&$ 0.152 $&$ 0.124 $& &$ 0.055 $&$ 0.100 $&$ 0.143 $&$ 0.143 $& &$ 0.072 $&$ 0.029 $&$ 0.082 $&$ 0.082 $& &$ 0.032 $&$ 0.067 $&$ 0.144 $&$ 0.144$  \\ 
 %&  &  &  &  & &  &  &  &  &  &  &  &  &  &  &  & & \\ 
 %\cline{2-19}	  
&                              
  \multirow{2}{*}{\centering $30$}
	  & $10$ & $0.508 $&$ 0.210 $&$ 0.186 $&$ 0.165 $& &$ 0.259 $&$ 0.036 $&$ 0.142 $&$ 0.142 $& &$ 0.249 $&$ 0.038 $&$ 0.082 $&$ 0.082 $& &$ 0.353 $&$ 0.038 $&$ 0.142 $&$ 0.142$  \\ 
	  &  & $30$ & $0.127 $&$ 0.270 $&$ 0.168 $&$ 0.166 $& &$ 0.055 $&$ 0.100 $&$ 0.157 $&$ 0.157 $&&$ 0.074 $&$ 0.057 $&$ 0.091 $&$ 0.091 $& &$ 0.032 $&$ 0.067 $&$ 0.160 $&$ 0.160 $ \B\\ 
   \hline
		\end{tabular}
}
\label{tab:LTbetas}
\end{table}

\begin{table}[htbp!]
\centering
\caption{Log-$t$-Student promotion time model: Estimates of $\eta$ and $\phi$ (with true values  $\eta = 5$ and $\phi = 1$).}
		\begin{tabular}{ccccccccccccc}
			\hline
\multirow{2}{*}{\centering $n$} & \multirow{2}{*}{\centering $cp$} & \multirow{2}{*}{\centering $cf$} & \multicolumn{4}{c}{$\widehat{\eta}$}& &\multicolumn{4}{c}{$\widehat{\phi}$}\Top\B\\
 \cline{4-7} \cline{9-12}
 			& & & mean &  RB & $\sqrt{\mbox{RSME}}$ & se && mean & RB & $\sqrt{\mbox{RSME}}$ & se \Top\B \\
\hline 
\multirow{4}{*}{\centering 250}
 & 
 \multirow{2}{*}{\centering $15$}
	  & $10$ &  $4.948 $&$ -0.010 $&$ 0.820 $&$ 0.818 $& &$ 1.028 $&$ 0.028 $&$ 0.175 $&$ 0.173$ \Top\\ 
	  &  & $30$ &  $4.726 $&$ -0.055 $&$ 0.686 $&$ 0.628 $& &$ 1.016 $&$ 0.016 $&$ 0.203 $&$ 0.203$  \\ 
 %&  &  &  &  & &  &  &  &  &  &  &  &  &  &  &  & & \\ 
 %\cline{2-19}	  
&                        
  \multirow{2}{*}{\centering $30$}
	  & $10$ &  $6.512 $&$ 0.302 $&$ 3.145 $&$ 2.757 $& &$ 1.059 $&$ 0.059 $&$ 0.185 $&$ 0.176$ \\ 
	  &  & $30$ &  $6.691 $&$ 0.338 $&$ 3.770 $&$ 3.370 $& &$ 1.195 $&$ 0.195 $&$ 0.323 $&$ 0.258 $\\ 
   &  &  &  &  &  &  &  &  &  &    \\
\multirow{4}{*}{\centering 500}
 &
 \multirow{2}{*}{\centering $15$}
	  & $10$ &  $4.883 $&$ -0.023 $&$ 0.567 $&$ 0.555 $& &$ 1.029 $&$ 0.029 $&$ 0.127 $&$ 0.124$ \\ 
	  &  & $30$ &  $4.677 $&$ -0.065 $&$ 0.533 $&$ 0.424 $& &$ 1.014 $&$ 0.014 $&$ 0.141 $&$ 0.140$  \\  
 %\cline{2-19}	  
&                              
  \multirow{2}{*}{\centering $30$}
	  & $10$ & $ 6.196 $&$ 0.239 $&$ 1.915 $& $ 1.495 $& &$ 1.076 $&$ 0.076 $&$ 0.148 $&$ 0.127$ \\ 
	  &  & $30$ & $6.164 $&$ 0.233 $&$ 2.018 $&$ 1.649 $& &$ 1.204 $&$ 0.204 $&$ 0.276 $&$ 0.186$  \\ 
   &  &  &  &  &  &  &  &  &  &   \\
\multirow{4}{*}{\centering 1000}
 &
 \multirow{2}{*}{\centering $15$}
	  & $10$    &  $4.848 $&$ -0.030 $&$ 0.409 $&$ 0.380 $& &$ 1.030 $&$ 0.030 $&$ 0.092 $&$ 0.087$ \\ 
	  &  & $30$ & $4.654 $&$ -0.069 $&$ 0.457 $&$ 0.298 $& &$ 1.012 $&$ 0.012 $&$ 0.100 $&$ 0.099$ \\ 
 %&  &  &  &  & &  &  &  &  &  &  &  &  &  &  &  & & \\ 
 %\cline{2-19}	  
&                              
  \multirow{2}{*}{\centering $30$}
	  & $10$    & $6.041 $&$ 0.208 $&$ 1.433 $&$ 0.984 $& &$ 1.083 $&$ 0.083 $&$ 0.123 $&$ 0.091$ \\ 
	  &  & $30$ & $5.996 $&$ 0.199 $&$ 1.462 $&$ 1.070 $& &$ 1.207 $&$ 0.207 $&$ 0.248 $&$ 0.136 $ \B\\ 
   \hline
\end{tabular}
\label{tab:LTetaphi}
\end{table}

%%%%%%%%%%%%%%%%%%%%%%%%%%%%%%%%%%%%%%%%%%%%%%%%%%%%%%%%%%%%%%%%%%%%%%%%%%%%%%%%%%%%%%%%%%%%%%%%%%%%%
%                           MODELO BS
%%%%%%%%%%%%%%%%%%%%%%%%%%%%%%%%%%%%%%%%%%%%%%%%%%%%%%%%%%%%%%%%%%%%%%%%%%%%%%%%%%%%%%%%%%%%%%%%%%%%%

%\newpage
%\begin{landscape}
\begin{table}[htbp!]
\caption{Birnbaum-Saunders promotion time model: Estimates of the regression coefficients  for different values of $n$, censoring percentage ($cp$), and cure fraction ($cf$), with true values of $\vecbeta$ for each  $cf$ $(\vecbeta_{cf})$ given by   $\vecbeta_{10}= (0.42,0.25,0.24,0.34)^\top$ and $\vecbeta_{30} =(0.10,0.05,0.07,0.03)^\top$.}
%\resizebox{23cm}{!}{
\resizebox{\linewidth}{!}{ 
		\begin{tabular}{cccccccccccccccccccccc}
			\hline
\multirow{2}{*}{\centering $n$} & \multirow{2}{*}{\centering $cp$} & \multirow{2}{*}{\centering $cf$}& \multicolumn{4}{c}{$\widehat{\beta}_0$}& &\multicolumn{4}{c}{$\widehat{\beta}_1$}& & \multicolumn{4}{c}{$\widehat{\beta}_2$}& &\multicolumn{4}{c}{$\widehat{\beta}_3$}\Top\B\\
 \cline{4-7} \cline{9-12} \cline{14-17} \cline{19-22}   
 					  & & 	& mean &  RB & $\sqrt{\mbox{RSME}}$ & se & & mean &  RB & $\sqrt{\mbox{RSME}}$ & se & & mean &  RB & $\sqrt{\mbox{RSME}}$ & se && mean &  RB & $\sqrt{\mbox{RSME}}$ & se \Top\B\\
\hline 
\multirow{4}{*}{\centering 250}
 & 
 \multirow{2}{*}{\centering $15$}
	  & $10$    &  $0.353 $&$ -0.160 $&$ 0.237 $&$ 0.227 $& &$ 0.274 $&$ 0.096 $&$ 0.260 $&$ 0.259 $& &$ 0.258 $&$ 0.075 $&$ 0.151 $&$ 0.150 $& &$ 0.366 $&$ 0.076 $&$ 0.272 $&$ 0.271$ \Top\\ 
	  &  & $30$ & $-0.016 $&$ -1.160 $&$ 0.263 $&$ 0.236 $& &$ 0.053 $&$ 0.060 $&$ 0.293 $&$ 0.293 $& &$ 0.072 $&$ 0.029 $&$ 0.168 $&$ 0.168 $& &$ 0.030 $&$ 0.000 $&$ 0.302 $&$ 0.302$ \\  
 %&  &  &  &  & &  &  &  &  &  &  &  &  &  &  &  & & \\ 
 %\cline{2-19}	  
&                        
  \multirow{2}{*}{\centering $30$}
	   & $10$    &  $0.554 $&$ 0.319 $&$ 0.334 $&$ 0.306 $& &$ 0.272 $&$ 0.088 $&$ 0.279 $&$ 0.279 $& &$ 0.261 $&$ 0.088 $&$ 0.167 $&$ 0.166 $& &$ 0.363 $&$ 0.068 $&$ 0.297 $&$ 0.296$  \\ 
	  &  & $30$ & $0.243 $&$ 1.430 $&$ 0.345 $&$ 0.314 $& &$ 0.051 $&$ 0.020 $&$ 0.317 $&$ 0.317 $& &$ 0.076 $&$ 0.086 $&$ 0.188 $&$ 0.188 $& &$ 0.029 $&$ -0.033 $&$ 0.330 $&$ 0.330$ \\   
  &  &  &  &  & &  &  &  &  &  &  &  &  &  &  &  \\
\multirow{4}{*}{\centering 500}
 &
 \multirow{2}{*}{\centering $15$}
	  & $10$    & $0.345 $&$ -0.179 $&$ 0.180 $&$ 0.164 $ &&$ 0.265 $&$ 0.060 $&$ 0.178 $&$ 0.178 $& &$ 0.260 $&$ 0.083 $&$ 0.107 $&$ 0.105 $& &$ 0.363 $&$ 0.068 $&$ 0.191 $&$ 0.190$ \\ 
	  &  & $30$ & $-0.021 $&$ -1.210 $&$ 0.214 $&$ 0.177 $& &$ 0.051 $&$ 0.020 $&$ 0.203 $&$ 0.203 $& &$ 0.074 $&$ 0.057 $&$ 0.118 $&$ 0.118 $& &$ 0.035 $&$ 0.167 $&$ 0.219 $&$ 0.219$ \\ 
 %&  &  &  &  & &  &  &  &  &  &  &  &  &  &  &  & & \\ 
 %\cline{2-19}	  
&                              
  \multirow{2}{*}{\centering $30$}
	   & $10$    & $0.520 $&$ 0.238 $&$ 0.237 $&$ 0.214 $& &$ 0.270 $&$ 0.080 $&$ 0.196 $&$ 0.195 $& &$ 0.259 $&$ 0.079 $&$ 0.115 $&$ 0.114 $& &$ 0.366 $&$ 0.076 $&$ 0.206 $&$ 0.204$  \\ 
	  &  & $30$ & $0.233 $&$ 1.330 $&$ 0.260 $&$ 0.224 $& &$ 0.053 $&$ 0.060 $&$ 0.222 $&$ 0.222 $&&$ 0.073 $&$ 0.043 $&$ 0.131 $&$ 0.131 $& &$ 0.030 $&$ 0.000 $&$ 0.233 $&$ 0.233$ \\ 
    &  &  &  &  &  &  &  &  &  &  &  &  &  &  &  &  \\
\multirow{4}{*}{\centering 1000}
 &
 \multirow{2}{*}{\centering $15$}
	   & $10$    & $0.342 $&$ -0.186 $&$ 0.139 $&$ 0.115 $ &&$ 0.267 $&$ 0.068 $&$ 0.129 $&$ 0.128 $& &$ 0.256 $&$ 0.067 $&$ 0.076 $&$ 0.074 $& &$ 0.365 $&$ 0.074 $&$ 0.132 $&$ 0.130$ \\ 
	  &  & $30$ & $-0.021 $&$ -1.210 $&$ 0.172 $&$ 0.123 $& &$ 0.056 $&$ 0.120 $&$ 0.143 $&$ 0.143 $& &$ 0.073 $&$ 0.043 $&$ 0.083 $&$ 0.083 $& &$ 0.029 $&$ -0.033 $&$ 0.146 $&$ 0.146$ \\ 
 %&  &  &  &  & &  &  &  &  &  &  &  &  &  &  &  & & \\ 
 %\cline{2-19}	  
&                              
  \multirow{2}{*}{\centering $30$}
	   & $10$    & $0.514 $&$ 0.224 $&$ 0.175 $&$ 0.147 $& &$ 0.266 $&$ 0.064 $&$ 0.140 $&$ 0.139 $& &$ 0.258 $&$ 0.075 $&$ 0.084 $&$ 0.082 $& &$ 0.365 $&$ 0.074 $&$ 0.144 $&$ 0.142$ \\ 
	  &  & $30$ & $0.227 $&$ 1.270 $&$ 0.201 $&$ 0.155 $& &$ 0.052 $&$ 0.040 $&$ 0.155 $&$ 0.155 $&&$ 0.076 $&$ 0.086 $&$ 0.092 $&$ 0.092 $& &$ 0.033 $&$ 0.100 $&$ 0.159 $&$ 0.159$ \B\\ 
   \hline
		\end{tabular}
}
%}
\label{tab:BSbetas}
\end{table}

%\newpage
%\begin{landscape}
\begin{table}[htbp!]
\centering
%\scriptsize
\caption{Birnbaum-Saunders promotion time model: Estimates of $\eta$ and $\phi$ (with true values  $\eta = 5$ and $\phi = 1$).}
%\resizebox{23cm}{!}{
%\resizebox{\linewidth}{!}{ 
		\begin{tabular}{cccccccccccc}
			\hline
\multirow{2}{*}{\centering $n$} & \multirow{2}{*}{\centering $cp$} & \multirow{2}{*}{\centering $cf$} & \multicolumn{4}{c}{$\widehat{\eta}$}& &\multicolumn{4}{c}{$\widehat{\phi}$}\Top\B\\
 \cline{4-7} \cline{9-12}
 			& & & mean &  RB & $\sqrt{\mbox{RSME}}$ & se && mean &  RB & $\sqrt{\mbox{RSME}}$ & se \Top\B \\
\hline 
\multirow{4}{*}{\centering 250}
 & 
 \multirow{2}{*}{\centering $15$}
	  & $10$    & $5.122 $&$ 0.024 $&$ 0.657 $&$ 0.646 $& &$ 1.048 $&$ 0.048 $&$ 0.196 $&$ 0.191$ \Top\\ 
	  &  & $30$ & $4.848 $&$ -0.030 $&$ 0.424 $&$ 0.396 $& &$ 0.986 $&$ -0.014 $&$ 0.141 $&$ 0.140$ \\ 
 %&  &  &  &  & &  &  &  &  &  &  &  &  &  &  &  & & \\ 
 %\cline{2-19}	  
&                        
  \multirow{2}{*}{\centering $30$}
	   & $10$    & $6.320 $&$ 0.264 $&$ 2.363 $&$ 1.959 $& &$ 1.324 $&$ 0.324 $&$ 0.513 $&$ 0.397$ \\ 
	  &  & $30$ &  $6.783 $&$ 0.357 $&$ 2.906 $&$ 2.294 $& &$ 1.482 $&$ 0.482 $&$ 0.659 $&$ 0.450$ \\  
   &  &  &  &  &  &  &  &  &  &    \\
\multirow{4}{*}{\centering 500}
 &
 \multirow{2}{*}{\centering $15$}
	   & $10$    & $5.049 $&$ 0.010 $&$ 0.415 $&$ 0.413 $& &$ 1.036 $&$ 0.036 $&$ 0.131 $&$ 0.126$ \\ 
	  &  & $30$ & $4.823 $&$ -0.035 $&$ 0.321 $&$ 0.267 $& &$ 0.983 $&$ -0.017 $&$ 0.098 $&$ 0.097$\\    
 %\cline{2-19}	  
&                              
  \multirow{2}{*}{\centering $30$}
	   & $10$    & $5.992 $&$ 0.198 $&$ 1.390 $&$ 0.974 $& &$ 1.266 $&$ 0.266 $&$ 0.356 $&$ 0.237$ \\ 
	  &  & $30$ & $6.518 $&$ 0.304 $&$ 1.856 $&$ 1.069 $& &$ 1.444 $&$ 0.444 $&$ 0.516 $&$ 0.264$ \\   
   &  &  &  &  &  &  &  &  &  &   \\
\multirow{4}{*}{\centering 1000}
 &
 \multirow{2}{*}{\centering $15$}
	   & $10$    & $5.027 $&$ 0.005 $&$ 0.283 $&$ 0.282 $& &$ 1.032 $&$ 0.032 $&$ 0.093 $&$ 0.087$  \\ 
	  &  & $30$ & $4.809 $&$ -0.038 $&$ 0.264 $&$ 0.182 $& &$ 0.982 $&$ -0.018 $&$ 0.068 $&$ 0.065$ \\ 

 %\cline{2-19}	  
&                              
  \multirow{2}{*}{\centering $30$}
	  & $10$    & $5.886 $&$ 0.177 $&$ 1.080 $&$ 0.617 $& &$ 1.245 $&$ 0.245 $&$ 0.293 $&$ 0.159$ \\ 
	  &  & $30$ & $6.431 $&$ 0.286 $&$ 1.575 $&$ 0.656 $& &$ 1.431 $&$ 0.431 $&$ 0.465 $&$ 0.173$ \B\\ 
   \hline
		\end{tabular}
%}
%}
\label{tab:BSetaphi}
\end{table}

%%%%%%%%%%%%%%%%%%%%%%%%%%%%%%%%%%%%%%%%%%%%%%%%%%%%%%%%%%%%%%%%%%%%%%%%%%%%%%%%%%%%%%%%%%%%%%%%%%%%%%%%%%%%%%%%%%%%%%%%%%%%%%%%%%%%%%%%%%%%
%                                     Applications
%%%%%%%%%%%%%%%%%%%%%%%%%%%%%%%%%%%%%%%%%%%%%%%%%%%%%%%%%%%%%%%%%%%%%%%%%%%%%%%%%%%%%%%%%%%%%%%%%%%%%%%%%%%%%%%%%%%%%%%%%%%%%%%%%%%%%%%%%%%%%
%\newpage
\section{Empirical application} \label{MRLS-sec:5}
\noindent

To illustrate the applicability of the proposed log-symmetric model with cure fraction, we considered data on leprosy patients. Leprosy is a chronic and contagious disease with slow evolution and a high degree of disability. Some leprosy patients have reactive states or leprosy reactions. These reactions are the main causes of patients' physical disabilities and deformities,  but they may not occur for some patients.
	
The dataset refers to a retrospective study conducted between 2010 and 2014 at the  Institute of Tropical Medicine (IMT) of the Universidade Federal do  Rio Grande do Norte, Brazil. The medical records of 263 patients diagnosed with leprosy were evaluated. For each patient, the lifetime corresponds to the time (in months) between the disease diagnosis and the first leprosy reaction. For  $44\%$ of the patients, the leprosy reaction was not observed, which corresponds to the censoring proportion. %We note that the median  of censored times was near to  47 months, while the median of time to leprosy was approximately 7 months.

\begin{figure}[htbp]
\center
\includegraphics[scale=0.5]{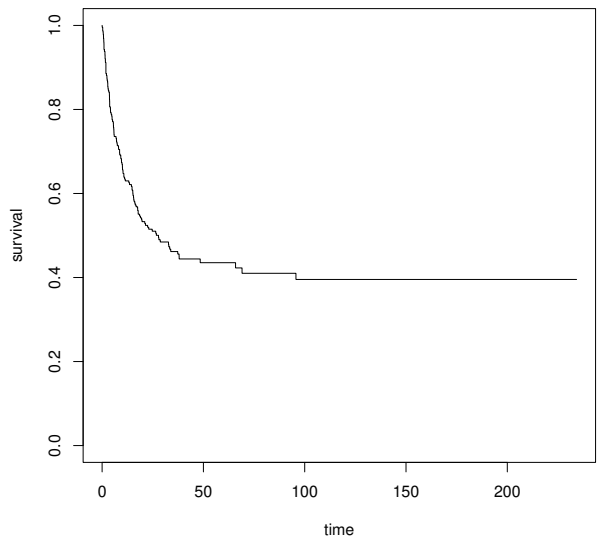}
\caption{Empirical survivor function (Kaplan-Meier) for the data.}
\label{img6}
%\small{Fonte: Elaborada pelo próprio autor.}
\end{figure}

In Figure \ref{img6}, we present the Kaplan-Meier curve of the observed data. There are indications of the use of the survival cure rate model since  the survival curve does not tend to zero in a sufficiently long follow-up time but stabilizes around $40\%$.

Initially, we did not consider covariates in the regression structure and fitted the following cure rate models: standard mixture, promotion time, and geometric models, considering the Weibull, log-normal, log-$t$-Student, and Birnbaum-Saunders distributions for the latency. Although the Weibull distribution does not belong to the log-symmetric class of distributions, it  was used for comparison. In terms of model selection criteria, we can use the Akaike information criterion (AIC) and the Bayesian information criterion (BIC) \citep{BurnhamAnderson2004}. According to Table \ref{tabmod1}, the log-$t$-Student standard mixture model with $\nu = 8$ presented the smallest values of AIC and BIC, thus being the best distribution fitted to these data.

\begin{table}[]
\begin{center}
\caption{Values of AIC and BIC for the fitted distributions.}
\label{tabmod1}
\begin{tabular}{ccccc}
\hline
Incidence distribution    & Latency distribution  & Add parameter & AIC   & BIC\Top\B   \\
\hline
\multirow{10}{*}{Bernoulli} & Weibull                        & -   & 1196.727 & 1267.526\Top\B\\
\cline{2-5}
                            & log normal                     & -   & 1194.764 & 1265.563\Top\B\\
\cline{2-5}
                          & \multirow{4}{*}{log-$t$-student} & 2   & 1201.386 & 1272.185\Top\\
                            &                                & 4   & 1194.542 & 1265.341\\
                            &                                & 6   & 1193.161 & 1263.960\\
                            &                                & 8   & \bf{1192.809} & \bf{1263.608}\B\\
\cline{2-5}
                            & \multirow{4}{*}{Birnbaum-Saunders}& 1.2  & 1207.747 & 1278.546\Top \\
                            &                                & 2    & 1215.362 & 1286.161 \\
                            &                                & 2.8  & 1224.626 & 1295.425\\
                            &                                & 3.6  & 1233.847 & 1304.646\B\\
\hline
\multirow{10}{*}{Poisson}  & Weibull                        & -     & 1195.096 & 1265.896\Top\B\\
\cline{2-5}
                            & log normal                     & -    & 1195.034 & 1265.833\Top\B\\
\cline{2-5}
                            & \multirow{4}{*}{log-$t$-student} & 2  & 1203.658 & 1274.457\Top\\
                            &                                & 4    & 1195.095 & 1265.894 \\
                            &                                & 6    & 1193.396 & 1264.195 \\
                            &                                & 8    & 1192.960 & 1263.760\B\\
\cline{2-5}
                            & \multirow{4}{*}{Birnbaum-Saunders} & 1.2  & 1203.082 & 1273.881\Top\B\\
                            &                                & 2    & 1203.003 & 1273.802 \\
                            &                                & 2.8  & 1202.687 & 1273.486 \\
                            &                                & 3.6  & 1202.526 & 1273.325\B\\
\hline
\multirow{10}{*}{Geometric} & Weibull                        & -    & 1193.831 & 1264.630\Top\B\\
\cline{2-5}
                            & log normal                     & -    & 1195.399 & 1266.198\Top\B\\
\cline{2-5}
                            & \multirow{4}{*}{log-$t$-student} & 2  & 1205.927 & 1276.726\Top\\
                            &                                & 4    & 1195.498 & 1266.297 \\
                            &                                & 6    & 1193.555 & 1264.355 \\
                            &                                & 8    & 1193.098 & 1263.897\B \\
\cline{2-5}
                            & \multirow{4}{*}{Birnbaum-Saunders}& 1.2   & 1199.522 & 1270.321\Top\\
                            &                                & 2     & 1199.443 & 1270.243 \\
                            &                                & 2.8   & 1199.378 & 1270.178 \\
                            &                                & 3.6   & 1199.347 & 1270.146\B \\
\hline
%\multicolumn{5}{l}{{\small $^a$ $\nu$ and $\alpha$ represent the additional parameters of the log-$t$-Student and Birnbaum-Saunders distributions, respectively}.}
\end{tabular}
\end{center}
\end{table}

\subsection{Including covariates}
\noindent

\begin{figure}
\center
\subfloat[][\textit{gender}]{\includegraphics[width=6cm]{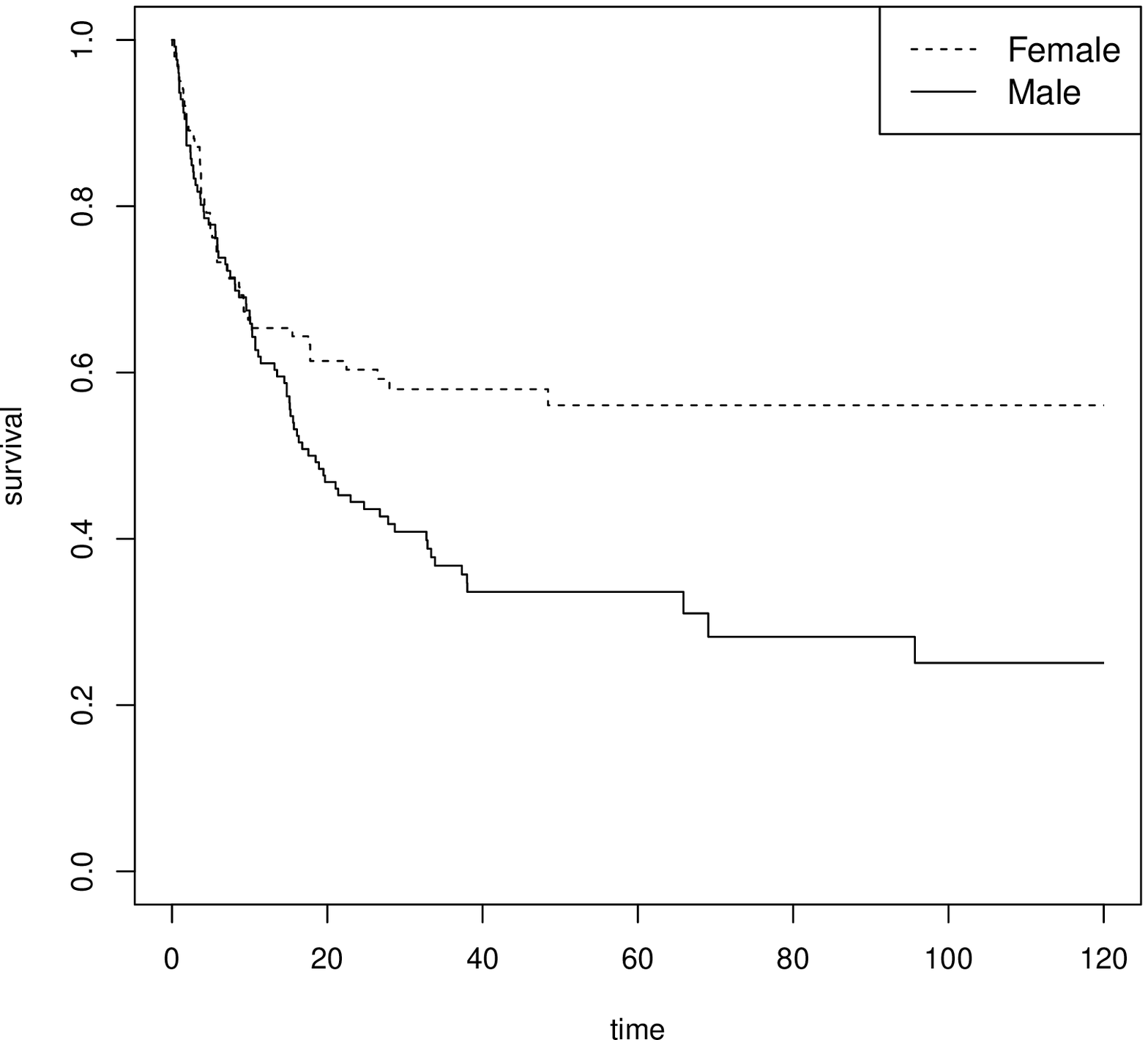}\label{fig:gender}} \hspace{2cm}
\subfloat[][\textit{Leprosy Classification}]{\includegraphics[width=6cm]{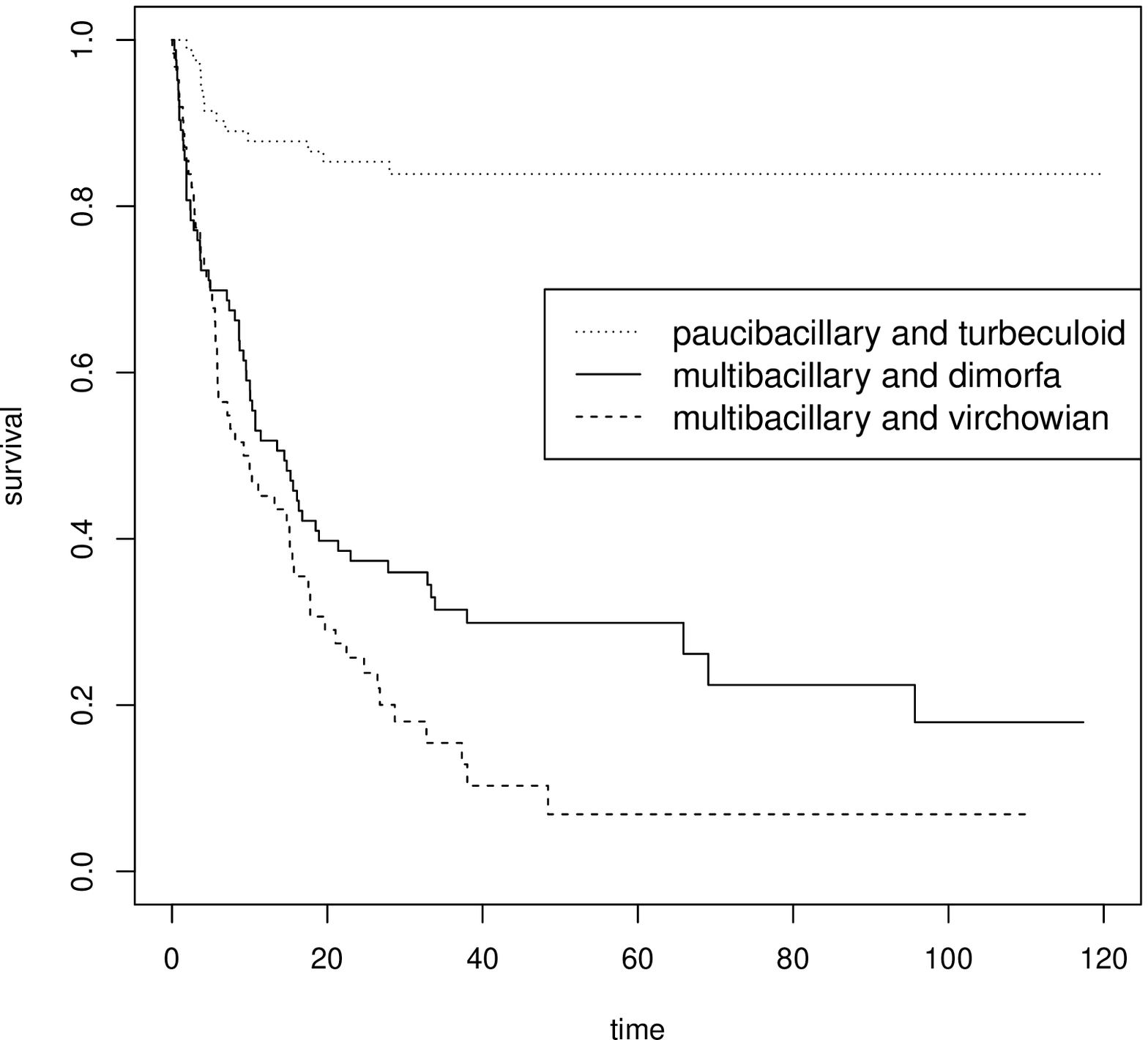}\label{fig:LC}}
\caption{\small Empirical survivor function (Kaplan-Meier) for \textit{gender} and \textit{Leprosy Classification}.}
\end{figure}

To evaluate the  effect of factors on the proportion of individuals immune to leprosy reactions, we considered the variables \textit {gender}, \textit {age}, and \textit{Leprosy Classification}. The  Leprosy Classification (LC) in  patients  was according to the clinical form of the disease (Tuberculoid, Dimorfa, or Virchowiana) and the operational classification (Paucibacillary - cases with up to 5 lesions and Multubacillary - cases with over 5 lesions). Thus, in this factor, patients could only be classified as Paucibacillary and Turbeculoid, Multibacillary and Dimorfa, and Multibacillary and Virchowian.
In Figures \ref{fig:gender} and \ref{fig:LC}, we present the Kaplan-Meier curves  according to the categorical variables \textit{gender} and \textit {Leprosy Classification}, respectively.

%%%%%%%%%%%%%%%%%%%%%% Mudar graficos: female   male
The log-$t$-Student standard mixture model with $\nu = 8$ was fitted to the data. %, considering the covariables \textit{gender, age} and \textit{Leprosy classification} in order to identify the covariates that better explain the  immune rate. % or " the proportion of patients immune to leprosy reactions."
Considering likelihood ratio tests, the final model contains only the factor LC. %(see Table \ref{tabfactor}).
The results of the final model are presented in Table \ref{tabmodreg}.

%\begin{table}[ht]
%\centering
%\caption{Fitted log-$t$-Student standard mixture model.}%  and tests of  hypothesis} \small{ $H_0:\beta_j=0$ versus $H_1:\beta_j \neq 0$, $j=0,1,2$.}}
%\label{tabmodreg}
%\begin{tabular}{ccc}
%\hline 
%\multirow{2}{*}{Factor}&\multirow{2}{*}{\centering Parameters} &\multirow{2}{*}{\centering Estimates (SE)}\\
%    & &   \\ \hline
%intercept & $\beta_{0}$ & $-1.551 (0.429)$  \\ \hline
%\multirow{2}{*} 
% & $\beta_{1}$ & $-3.264 (6.045)$  \\ 
%LC & $\beta_{2}$ & $3.063 (0.515)$  \\ \hline
% median &$\eta$ & $8.787(0.324)$  \\ \hline
%asymmetry &$\phi$ & $1.862(0.183)$ 
% \\\hline
%\end{tabular}
%\label{tab3} \\
%%\small{Fonte: Elaborada pelo próprio autor.}
%\end{table}

\begin{table}[]
\begin{center}
\caption{Fitted log-$t$-Student standard mixture model.}
\label{tabmodreg}
\begin{tabular}{lcc}
\hline
Factor              & Parameter  & Estimate (se) \Top\B\\
\hline
Intercept           & $\beta_0$   & $-$1.551 (0.429)  \Top\B\\
\hline
\multirow{2}{*}{LC} & $\beta_1$   & $-$3.264 (6.045)  \Top\\
                    & $\beta_2$   & 3.063 (0.515)   \\
\hline
Median              & $\eta$      & 8.787 (0.324)   \Top\B\\
\hline
Asymmetry           & $\phi$      & 1.862 (0.183)  \Top\B\\
\hline
\end{tabular}
\end{center}
\end{table}

%\newpage
We used the fitted model to estimate the immune fraction  for each level of the factor LC according to %\ref{tabfactorfc}.
\begin{equation*}
p_{\widehat{\theta_{i}}}(0)=\frac{\exp(-1.551-3.264x_{i1}+3.063x_{i2})}{1 +\exp(-1.551-3.264x_{i1}+3.063x_{i2})}. \nonumber
\end{equation*}

\begin{figure}[htbp]
\center
\includegraphics[scale=0.7]{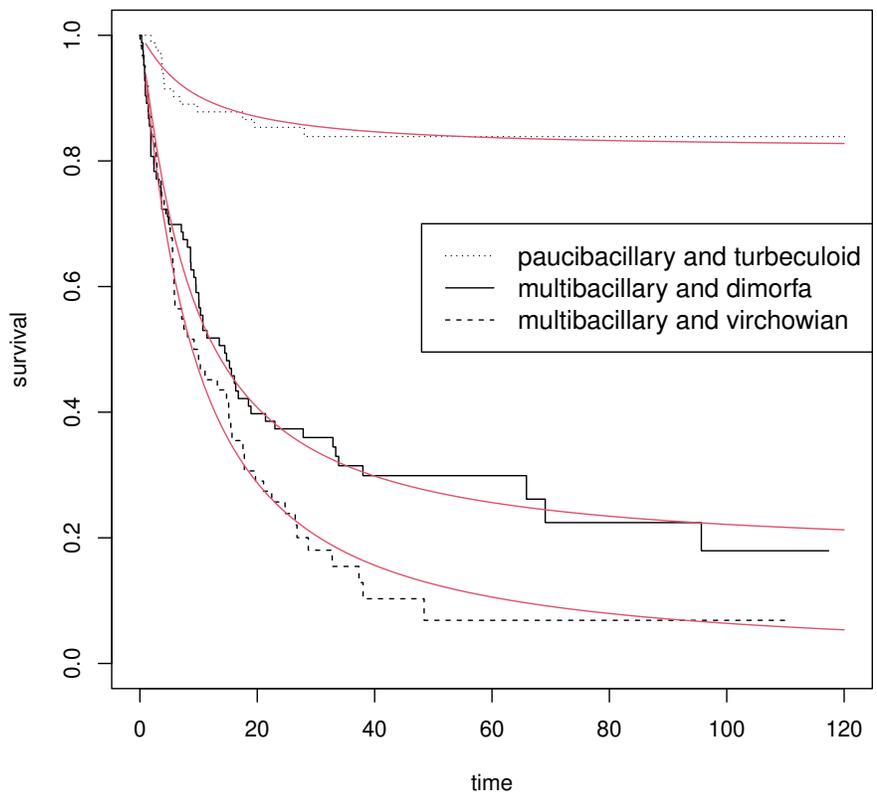}
\caption{Empirical survivor function (Kaplan-Meier) for the data according to Leprosy Classification and curves fitted by final model.}
\label{fit}
%\small{Fonte: Elaborada pelo próprio autor.}
\end{figure}

The results are summarized in Table \ref{tabfactorfc}. We can see that the immune fraction of  patients with Multibacillary and Dimorphic leprosy classification  was $ 17.5 \% $, whereas patients with Multibacillary and Virchowian classification  (the most severe) had a cure fraction  of $ 0.8 \% $, that is, there are practically no immune patients in this classification. Patients with leprosy classification as Paucibacillary and Tuberculoid (the  early stage  of the disease) had little chance of having these reactions since the estimated immune fraction was $ 82\% $. The estimated median time until the leprosy reaction of susceptible individuals was $\widehat{\eta} = 8.79 $ months, which is close to the empirical  median time ($7.37$). Figure \ref{fit}  indicates a good fit of the final model. % of the susceptible individuals ($7.37$). %see Figure \ref{img6}). 
%Table \ref{tabfactorfc} summarizes the results.
%(\ref{eq:fc})

% Tentando juntar as tabelas 9 e 11

\begin{table}[!h]
\begin{center}
\caption{Estimated immune fraction to leprosy reaction for levels of the LC factor with log-$t$-Student standard mixture model fitted to the data.}
\label{tabfactorfc}
\begin{tabular}{lccc}
\hline\multirow{2}{*}{Leprosy Classification} &\multicolumn{2}{c}{Indicator variables} \Top\B\\
\cline{2-3}& $x_1$ & $x_2$  & \multicolumn{1}{c}{cure fraction} \\\hline
Multibacilar and Dimorfa & 0 & 0 & \multicolumn{1}{c}{$17.5\%$}\Top\\ 
Multibacilar and Virchowiana & 1 & 0 & \multicolumn{1}{c}{$0.8\%$}\\
Paucibacilar and Tuberculoide & 0 & 1 & \multicolumn{1}{c}{$82.0\%$}\B\\ \hline
\end{tabular} 
%\small{Fonte: Elaborada pelo próprio autor.}
\end{center}
\end{table}
	
	% or: %we note that under this model the estimated median,

%%%%%%%%%%%%%%%%%%%%%%%%%%%%%%%%%%%%%%%%%%%%%%%%%%%%%%%%%%%%%%%%%%%%%%%%%%%%%%%%%%%%%%%%%%%%%%%%%%%%%%%%%%%%%%%%%%%%%%%%%%%%%%%%%%%%%%%%%%%%
%                                 Final remarks
%%%%%%%%%%%%%%%%%%%%%%%%%%%%%%%%%%%%%%%%%%%%%%%%%%%%%%%%%%%%%%%%%%%%%%%%%%%%%%%%%%%%%%%%%%%%%%%%%%%%%%%%%%%%%%%%%%%%%%%%%%%%%%%%%%%%%%%%%%%%%

\section{Final remarks}\label{MRLS-sec:6} 
\noindent

 In this paper, we proposed the long-term log-symmetric model. We considered a cure rate model in which the  {latency distribution}  belongs to the class of log-symmetric distributions. For the incidence, we considered the Bernoulli, Poisson, and geometric distributions. Covariates were included only in the parameter of the incidence distribution. 

We evaluated the performance of the maximum likelihood estimators of the model for some special cases of the log-symmetric promotion time  model through extensive Monte Carlo simulation studies. In general, the  bias and variability  of the maximum likelihood estimators increase as the censoring proportion and cure fraction increase, but they decrease as the sample size increases. We noted that  the  estimates for the intercept, $\eta$, and $\phi$ are more affected in the presence of censoring and cure fraction than the estimates of the regression coefficients.

In the empirical application, we verified that the log-$t$-Student standard mixture model presented the best fit to the data on time to leprosy reaction. The model could identify that the proportion of individuals immune to reactions differs with respect to the classification of leprosy, providing estimates to the  proportion of immunity according to the classification. Moreover, the proposed model also provided a general estimate of the median time  until the  reaction for susceptible individuals.

Considering the importance of a correct choice of the latency distribution and obtaining adequate estimates for the cure fraction \cite{yu2004}, we believe that this new class is helpful since it allows for adjusting several latency distributions.  This was illustrated in the application,  where we fitted different models to the data, including four distributions for latency and three for incidence, and according to AIC and BIC criteria, we could choose the best one. %to provide an adequate choice
 
In future work, we envisage to extend this proposed class  to accommodate covariates in the latency part (suitably incorporated in the $\eta$ (median) parameter) to allow for separate interpretation of the effects of the covariates on the cure fraction and failure time distribution of the uncured. In the simulation studies,  we noted an influence of the choice of the value representing the length of the follow-up on the estimation of the intercept in the incidence distribution. \cite{yu2004} have studied this in a mixture model context, so extending this study to the log-symmetric promotion time and geometric models is another interesting topic for future work.

%We evaluated the performance of the maximum likelihood estimates of the parameters of some special cases of the proposed model through extensive simulation studies. As result we note that in general, the variability  of the estimators  decreases as sample size grows and that the relative bias for the estimated coeficientes are  close to zero. However we note a high bias in the maximum likelihood estimates for the intercept ($ \beta_0 $) and in parameters, $ \eta $ and $ \phi $, in the presence of censorship and cure fraction, even in large samples.

%{\color{red} A relevant topic for future work is a comprehensive numerical comparison of the proposed model with other existing models and incorporate covariates on the parametric of the latency distribution. }

%%%%%%%%%%%%%%%%%%%%%%%%%%%%%%%%%%%%%%%%%%%%%%%%%%%%%%%%%%%%%%%%%%%%%%%%%%%%%%%%%%%%%%%%%%%%%%%%%%%%%%%%%%%%%%%%%%%%%%%%%%%%%%%%%%%%%%%%%%%%
%                                 Acknowledgments
%%%%%%%%%%%%%%%%%%%%%%%%%%%%%%%%%%%%%%%%%%%%%%%%%%%%%%%%%%%%%%%%%%%%%%%%%%%%%%%%%%%%%%%%%%%%%%%%%%%%%%%%%%%%%%%%%%%%%%%%%%%%%%%%%%%%%%%%%%%%%

\vspace{0.5cm}
%\section*{Acknowledgments}

%We gratefully acknowledge grants from the Brazilian agencie CAPES. %We thank the reviewers and the Associate Editor for their comments and suggestions.

%%%%%%%%%%%%%%%%%%%%%%%%%%%%%%%%%%%%%%%%%%%%%%%%%%%%%%%%%%%%%%%%%%%%%%%%%%%%%%%%%%%%%%%%%%%%%%%%%%%%%%%%%%%%%%%%%%%%%%%%%%%%%%%%%%%%%%%%%%%%
%                                          Appendix
%%%%%%%%%%%%%%%%%%%%%%%%%%%%%%%%%%%%%%%%%%%%%%%%%%%%%%%%%%%%%%%%%%%%%%%%%%%%%%%%%%%%%%%%%%%%%%%%%%%%%%%%%%%%%%%%%%%%%%%%%%%%%%%%%%%%%%%%%%%%%

\end{document}